\documentclass[journal]{IEEEtran}
\usepackage{color}
\usepackage{multirow}
\usepackage{cite}
\usepackage{balance}
\usepackage{amsmath}
\usepackage{enumitem}
\setlist[itemize]{noitemsep, topsep=0pt}

\usepackage{pdfsync}
\usepackage{amssymb}
\usepackage{bbold}
\usepackage{color}

\ifCLASSINFOpdf
\usepackage[pdftex]{graphicx}
\DeclareGraphicsExtensions{.pdf,.jpeg,.png, .jpg}
\else
\usepackage[dvips]{graphicx}

\DeclareGraphicsExtensions{.eps}
\fi
\graphicspath{{figs/}}
\usepackage{epstopdf}
\usepackage{booktabs}
\usepackage{amsmath}

\begin{document}

\title{Conservation Voltage Reduction (CVR) via Two-Timescale Control in Unbalanced Power Distribution Systems}

\author{Rahul Ranjan Jha,~\IEEEmembership{Student Member,~IEEE,}
		Anamika Dubey,~\IEEEmembership{Member,~IEEE,}
        		and  Kevin P. Schneider,~\IEEEmembership{Senior Member,~IEEE}
      \thanks{R. Jha, and A. Dubey,  are with the School of Electrical Engineering and Computer Science, Washington State University, Pullman, WA, 99164 e-mail: rahul.jha@wsu.edu,anamika.dubey@wsu.edu}
        \thanks{K. P. Schneider is with the Pacific Northwest National Laboratory, Battelle Seattle Research Center, Seattle, WA 98109 USA e-mail: kevin.schneider@pnnl.gov.}
}

\maketitle

\begin{abstract}
Voltage control devices are employed in power distribution systems to reduce the power consumption by operating the system closer to the lower acceptable voltage limits; this technique is called conservation voltage reduction (CVR). The different modes of operation for system's legacy devices (with binary control) and new devices (e.g. smart inverters with continuous control) coupled with variable photovoltaic (PV) generation results in voltage fluctuations which makes it challenging to achieve CVR objective. This paper presents a two-timescale control of feeder's voltage control devices to achieve CVR that includes (1) a centralized controller operating in a slower time-scale to coordinate voltage control devices across the feeder and (2) local controllers operating in a faster timescale to mitigate voltage fluctuations due to PV variability. The centralized controller utilizes a three-phase optimal power flow model to obtain the decision variables for both legacy devices and smart inverters. The local controllers operate smart inverters to minimize voltage fluctuations and restore nodal voltages to their reference values by adjusting the reactive power support. The proposed approach is validated using the IEEE-123 bus (medium-size) and R3-12.47-2 (large-size) feeders. It is demonstrated that the proposed approach is effective in achieving the CVR objective for unbalanced distribution systems.
\end{abstract}


\section{Introduction}
\label{Introduction}
\IEEEPARstart{D}{istribution} Driven by the requirement for improved energy efficiency, conservation voltage reduction (CVR) has emerged as a popular approach to reduce the power consumption of voltage dependent loads by operating the feeder at the lower limits of the acceptable voltages. It is shown in  \cite{schneider2010evaluation} that there is a 3.04\% of annual energy reduction by implementing CVR on all the distribution feeders throughout the United States. Traditionally, voltages  in electric power distribution systems are maintained by controlling the legacy voltage control devices such as capacitor banks, load tap changers and voltage regulators. With  IEEE 1547-2018 standard \cite{IEEE1547}, the distributed generators (DGs) equipped with smart inverters can help maintain distribution feeder voltages by providing reactive power support. To meet CVR objective, a centralized voltage control system can be envisioned that aims to minimize the feeder power consumption by coordinating all voltage control devices: capacitor switches, regulator taps, and DG smart inverters. Note that solar photovoltaic (PV) systems (both roof-top and utility scale) are the most prominent DG technology deployed in distribution feeders \cite{IEA}. 
Unfortunately, due to inherent variability of PV generation, the node voltages vary making it difficult for the centralized controller to maintain the desired nodal voltages to meet CVR objective.  To minimize the impacts of local PV variability on optimum decisions from centralized controller, there is a critical need for the local control of PV inverters that is coordinated with the centralized voltage control system. 

The related literature designs a centralized controller to achieve the CVR objective by solving a three-phase optimal power flow (OPF) problem to obtain decision variables for both legacy devices and smart inverters. The resulting OPF is a mixed integer nonlinear problem (MINLP) which is to be solved for three-phase unbalanced system and includes both discrete and continuous decision variables \cite{MINLPCVR,nojavan2014optimal,jabr2012minimum}. These methods, however, do not scale for a large-scale three-phase unbalanced power distribution systems. Moreover, the existing centralized control methods for CVR assume that the local generation do not vary during the discrete time intervals on which OPF decisions are implemented (typically 5-min or 15-min) \cite{paper1}. Unfortunately, this assumption is flawed as local PV generation is  significantly variable within a 5 or 15-min time-interval. The variation in local generation leads to nodal voltage fluctuations rendering the centralized decisions sub-optimal. The voltage fluctuations resulting from variable PV generation are also known to increase the number of operations for the legacy devices that may reduce their operating life. A local control of PV smart inverters can help mitigate the nodal voltage fluctuations caused by rapid changes in PV generation. However, the local control alone cannot help meet the CVR objectives that require coordination of all voltage control devices. {\em This calls for the coordination between the centralized and local controllers.} In this paper, we present a two-timescale control which coordinates the centralized and local controllers to achieve CVR objective in the presence of variable generation resources. 

The existing local voltage control methods are widely based on sensitivity matrix and volt-var droop characteristics of smart inverters \cite{shivashankar2016mitigating, Jensen, sensitivitybased, Zhao, Chamana, FarivarChe, EAbdelkarim, Liu,bokhari2016combined, zhang2018novel,zhu2016fast}. For the volt-var droop based method, the set points for the droop curve are predetermined to simply mitigate any over-voltage or under-voltage concerns \cite{shivashankar2016mitigating, sensitivitybased, Chamana}. These methods may cause voltages to oscillate, may lead to a steady state error, and are unable to optimize for system-wide objectives. The optimality and the stability of the droop based voltage control methods is discussed in \cite{zhu2016fast}. In order to remove the oscillations and improve stability, several authors \cite{Zhao,Chamana,FarivarChe} have proposed methods to dynamically change the droop points based on the local measurements. Dynamically obtaining droop set-points at shorter time-intervals (1-minute or less) remain a challenging problem. The sensitivity based approach which utilizes  $\frac{R}{X}$ ratio of the distribution system is used to obtain the required reactive power to reduce voltage fluctuations \cite{sensitivitybased}. However, here authors modelled the distribution system as an equivalent balanced single-phase system. In another work, the coefficients of change in active and reactive power are determined by repeatedly solving the power flow equations \cite{Liu}. The obtained coefficients, however, are not accurate for all operating conditions. In \cite{tanaka2010decentralised}, a decentralized approach for the voltage control of DGs is proposed by generating voltage reference for each inverters using PI control. In \cite{bonfiglio2014optimal}, authors propose a method based on feedback linearization where system voltage is maintained by providing the required reactive power to the PVs. Unfortunately, none of the above referred work, provides a mechanism for coordination among the centralized and local controllers; that is, the local controllers work independent of the centralized controller. 

Recently, a few researchers have proposed methods to coordinate the control of the centralized and local controllers \cite{garcia2014combined,bidgoli2018combined,weckx2014combined}. In these articles, the Q-V points for the volt-var droop curve are periodically updated by solving a centralized control problem. The local controls are used in sub-intervals to mitigate any sudden voltage violations. Although, these methods coordinate the  centralized and local control, they do not solve the problem for a three-phase unbalanced system and ignore the effect of control provided by other inverters in the system. In \cite{Andrew2017volt}, authors solve a three-phase OPF to obtain the volt-var curve for the smart inverter which can  perform autonomous voltage control; however, the obtained curve is based on a few specific planned  scenario and do not generalize for all possible operating conditions. Also, the volt-var curve required to maintain the pre-specified node voltages is generated without taking the effects of the var support from other inverters. The combined effect of CVR and power quality is solved in \cite{Ding} using the legacy devices and smart inverters. In this work, capacitor banks and voltage regulators are controlled heuristically to flatten the voltages to lower value while smart inverters work autonomously; thus, legacy devices and smart inverters are not really coordinated to meet the system-wide objectives. Thus, in the existing literature, the centralized and local control are not properly coordinated to meet CVR objective when local generation can fluctuate. 

Recently, several distributed algorithms have been proposed for the voltage and reactive power optimization in a power distribution system  \cite{almasalma2017dual,koukoula2015convergence,zheng2015fully,peng2016distributed}. For example, in \cite{almasalma2017dual,koukoula2015convergence}, a distributed voltage control approach based on dual decomposition is proposed for a balanced single-phase system. Here, the peer-to-peer communication is required to update the Lagrangian multipliers. The peer-to-peer communication increases the requirements for the communication infrastructure. In \cite{zheng2015fully}, the distribution system is separated into different areas to reduce the communication requirements compared to peer-to-peer communication. Here, the alternating direction method of multipliers (ADMM) is used to solve the distributed optimization problem for both an equivalent single-phase and three-phase unbalanced distribution system. The distributed algorithm still takes 100s of iterations and thus requires 100s rounds of communication among distributed agents to converge for a single step of the optimization problem. The  implementation of the existing distributed algorithms for the real-time voltage control will require a high-bandwidth and low-latency communication infrastructure. 
Thus, there is a critical need for low-compute and low-communication algorithms to optimize grid's voltage control devices while incorporating the effects of local generation variability. This paper presents a {\em two-timescale coordinated centralized and local control approach} for CVR. The CVR objective is achieved by optimally controlling feeder's legacy devices and smart inverters. First, a centralized controller is proposed that solves an OPF and obtains control set-points for feeder's voltage control devices for a three-phase unbalanced radial distribution system at discrete time intervals (every 5 to 15-min). Next, an adaptive local controller is proposed that cancels the effect of local generation variability on nodal voltage fluctuations and thus help maintain the same nodal voltages as obtained from the centralized controller based on the forecasted values of local generation.  
The overall contributions of this work are threefold: (1) we propose low-compute algorithms for the coordinated control of system's legacy devices and smart inverters for CVR; (2) we propose fast local control methods for smart inverters to mitigate the effects of local generation/load variability on nodal voltage fluctuations; (3) we propose low-compute and low-communication algorithms to coordinate the central and local controllers to achieve the CVR objectives for a system with significant levels of local generation variability. Note that the central controller needs to communicate with the local controllers at every 15-min time-interval, while the local controllers are operating at much faster time-intervals (real-time to every 1-min depending upon the granularity of the local measurements). 

The rest of the paper is organized as follows. In Section II, linear three-phase unbalanced power flow is presented. The mathematical models for the distribution system components are discussed in Section III. Section IV presents the proposed approach for the coordinated voltage control to achive the CVR objectives. Section V provides the mathematical formulation for the proposed local controllers.  The results and discussions are detailed in Section VI, followed by conclusion in Section VII. 

\section{Three Phase Unbalanced Power Flow}
In this section, we introduce the three-phase branch flow equations for an unbalanced power distribution system. Next, we detail a linearized approximate three-phase linear power flow model that is known to accurately approximate the nodal voltages. Note that achieving CVR objective requires a power flow model that accurately represents nodal voltages. For completeness, in the results section, we thoroughly validate the accuracy of the detailed three-phase linearized power flow model.   

 \subsection{Branch Flow Equations}
 The three-phase power flow equations as detailed in \cite{gan2014convex} are described in  (1)-(4). Let there be directed graph as shown in Fig.1, $\mathcal{G} = (\mathcal{N}, \mathcal{E})$ where $\mathcal{N}$ denotes set of buses and $\mathcal{E}$ denotes set of lines. Each line connects ordered pair of buses $(i,j)$ between two adjacent nodes $i$ and $j$. Let $\{a,b,c\}$ denotes the three phases of the system and $\Phi_i$ denotes set of phases on bus $i$. For each bus $i \in \mathcal{N}$, let phase $p$ complex voltage is given by $V_i^{p}$ and phase $p$ complex power demand is $s_{L,i}^p$. Let $V_i := [V_i^{p}]_{p \in \Phi_i}$ and $s_{L,i} := [s_{L,i}^{p}]_{p \in \Phi_i}$. For each line, let $p$ phase current be $I_{ij}^{p}$ and define, $I_{ij} := [I_{ij}^{p}]_{p \in \Phi_i}$. Let $z_{ij}$ be the phase impedance matrix.
\begin{eqnarray} \label{eq:1}
v_j  = v_i - (S_{ij}z_{ij}^H+z_{ij}S_{ij}^H) + z_{ij}l_{ij} z_{ij}^H \\
\text{diag}(S_{ij} - z_{ij}l_{ij}) =  \sum_{k:j \rightarrow k}{\text{diag}(S_{jk})}  + s_{L,j}  \\
\left[
  \begin{array}{cc}
    v_i & S_{ij} \\
    S_{ij}^H & l_{ij} \\
  \end{array}
\right] = \left[
  \begin{array}{c}
    V_i\\
    I_{ij}\\
  \end{array}
\right]
\left[
  \begin{array}{cc}
    V_i\\
    I_{ij}\\
  \end{array}
\right]^H \\
\left[
  \begin{array}{cc}
    v_i & S_{ij} \\
    S_{ij}^H & l_{ij} \\
  \end{array}
\right] : - \text{Rank-1 PSD Matrix}
\end{eqnarray}

Here, (1) represents voltage drop equation, (2) corresponds to power balance equation, (3) are variable definitions for power flow quantities, and (4) is a Rank-1 constraint that makes the associated optimization problem non-convex. 

 \begin{figure}[t]
\centering
\includegraphics[width=3.2in]{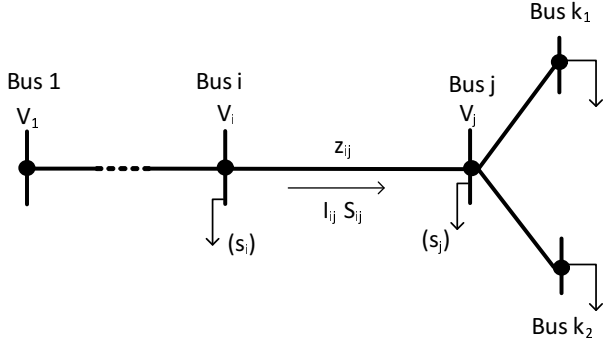}
\caption{\footnotesize Distribution system branch flow model.}
\label{fig:1}
\end{figure}

Although in the literature, researchers have proposed methods to to obtain a convex power flow formulation by either relaxing or  approximating the power flow equations \cite{Low1,Low2,gan2014convex}, there remain additional non-linearities due to load models and the nature of control decisions that require attention. For example, the commonly used ZIP load model that  combines constant impedance ($Z$), constant current ($I$)  and constant power ($P$) load models leads to additional non-linearities in the power flow model. Similarly, the decision variables for capacitor banks' switching and voltage regulators' taps, are discrete leading to a binary/mixed-integer problem. The resulting CVR optimization problem is thus a large-scale mixed integer nonlinear program (MINLP) that is computationally intensive to solve and hence not suitable for real-time control. These challenges led to the development of linearized three-phase power flow models that can realistically solve large-scale nonconvex OPF problems for unbalanced distribution systems with integer decision variables.  

\subsection{Linear Three-Phase Power Flow Model}
This section describes a three-phase linear power flow model that is an extension of single-phase distflow equations \cite{baran1989optimal}. The three-phase unbalanced power flow model involves nonlinear relationship between power, voltage and current in power balance and voltage equations as observed in equations (1) and (2). The additional nonlinearity results from the mutual coupling among the three phases. In linearized three-phase model, it is assumed that the power loss in a branch is relatively smaller as compared to the power flow in the branch. Due to this assumption,  the nonlinearity due to power loss and voltage drop  are ignored. Further, the nonlinearity due to mutual coupling is removed by approximating the voltage phase angle difference among the phases. It is assumed that the phase voltages are $\frac{2\pi}{3}$ apart \cite{gan2014convex}.
Mathematically, the approximation is represented as:
\begin{equation}\label{eq:5}
    \frac{V_i^a}{V_i^b} \simeq  \frac{V_i^b}{V_i^c} \simeq \frac{V_i^c}{V_i^a} = e^{j2\pi/3}=a^2
\end{equation}

Note that the apparent power can be represented using (6). We use the approximation (5) to reduce the number of variables in the optimization problem. Due to (5), the off diagonal terms in the power balance equation can be represented as a function of diagonal terms as shown in (7) \cite{gan2014convex}.
\begin{eqnarray}\label{eq:6}
    S^{aa}_{ij} = V^a_i*I^a_{ij} \hspace{.2 cm} \text{and} \hspace{.2 cm} S^{ab}_{ij} = V^a_i*I^b_{ij}\\
    S^{ab}_{ij} = V^a_i*I^b_{ij} = \frac{V^a_i}{V^b_i}*V^b_i*I^b_{ij} = a^2*S^{bb}
\end{eqnarray}

For the linear AC power flow, it is assumed that the branch power loss are relatively smaller as compared to the respective branch power flow. Hence, by ignoring the  power losses and using (1), (2) and (7), the linear AC branch flow equations are obtained (8)-(10), where,  $v_i^p = (V_i^p)^2$ and $v_j^p = (V_j^p)^2$. 

\begin{eqnarray} \label{eq:8}
    P^{pp}_{ij} &=& \sum_{k:j \rightarrow k} P^{pp}_{jk} + p^p_{Lj} \hspace{.7 cm}  p \in {a,b,c}\\
    Q^{pp}_{ij} &=& \sum_{k:j \rightarrow k} Q^{pp}_{jk} + q^p_{Lj} \hspace{.7 cm}  p \in {a,b,c}\\
   v_i^p - v_j^p &=& \sum_{q \in \phi_ j} 2\Re [S_{ij}^{pq}(z_{ij}^{pq})^*] \hspace{.4 cm}  p,q \in {a,b,c} 
\end{eqnarray}
   
Since the linear power flow equations do not include the components of power loss in active and reactive power flow equations, i.e (8) and (9), the line flows are not well approximated. Note that the voltage equation in (10) takes the effect of active and reactive power flows along the line on the feeder voltage drop; it however, does not include the effect of power losses along the lines on voltage drop. Since, the line losses are significantly less than the line flow due to load demand, the obtained voltage drops are well approximated. We further verify in Section 6 that the node voltages obtained using the linearized power flow model are well approximated for an unbalanced power distribution system. Since the CVR objective depends on nodal voltages, the AC linearized power flow equations are accurate for optimizing the CVR objective.  
 
\section{Distribution System Equipment Models}   
This section details the mathematical models for voltage dependent components of the distribution systems. As shown in Section 2, the approximate power flow equations are a function of the square of nodal voltage magnitude. Therefore, the distribution system components such as voltage regulators, capacitor banks, and customers loads are modeled as a function of the square of voltage magnitude. This ensures that the constraints for CVR optimization problem remain mixed-integer linear in the problem decision variables. These models were introduced in our prior work as well \cite{paper1}. 

\subsection{Voltage Regulators}  
The voltage regulation range of a voltage regulator is assumed to be $\pm10\%$, which is divided into 32-steps. The series and shunt impedance of the  voltage regulator have very small value \cite{kers} and hence these are ignored. Let, for a line $(i,j)$ the voltage regulator is connected to phase $p$ with the turn ratio of $a^p$. The $a^p$ has step change of 0.00625 pu for the values between 0.9 to 1.1. An additional node $i'$ is introduced to model the current equations. Let for $u_{tap,i}^p \in \{0,1\}$ be a binary variable defined for each regulator step position i.e. $i \in (1,2,...,32)$. Also, define a vector $b_i \in \{0.9, 0.90625 , ..., 1.1\}$. Then $V_i^p$, $V_j^p$ where $p \in \Phi_i \cap \Phi_j$ are given as follows:

\begin{eqnarray} \label{eq11}
V_j^p = V_{i'}^p = a^p  V_i^p 
\end{eqnarray}

where, $a^p = \sum\limits_{i=1}^{32} b_i u_{tap,i}^p$ and  $\sum\limits_{i=1}^{32} u_{tap,i}^p = 1$.\\

In order to express (11) as a function of $v_i^p = (V_i^p)^2$ and $v_j^p = (V_j^p)^2$, we take square of (11) and define $a_p^2 = A_p$ and $b_i^2 = B_i$. Further realizing that $(u_{tap,i}^p)^2 = u_{tap,i}^p$, (11) can be reformulated as (12). 

\begin{eqnarray} \label{eq12}
v_j^p = A^p \times v_i^p
\end{eqnarray}

 \subsection{Capacitor Banks}
The per-phase model for capacitor banks is developed. The reactive power generated by a capacitor bank, $q_{cap,i}^{p}$, is defined as a function of binary control variable $u_{cap,i}^p \in \{0,1\}$ indicating the status (On/Off) of the capacitor bank, its rated per-phase reactive power $q_{cap,i}^{rated,p}$, and the square of the bus voltage at bus $i$ for phase $p$, $v_{i}^p$.

\begin{equation} \label{eq13}
q_{cap,i}^{p} = u_{cap,i}^p q_{cap,i}^{rated,p} v_{i}^p
\end{equation}

The capacitor bank model is assumed to be voltage dependent and provides reactive power as a function of $v_{i}^p$ when connected, i.e. $u_{cap,i}=1$. For a three-phase capacitor bank, a common control variable, $u_{cap,i}^p$, is defined for each phase.

 \subsection{Distributed Generation with Smart Inverters}
The DGs are modeled as negative loads with a known active power generation equal to the forecasted value. The reactive power support from the DGs depends on the rating of the smart inverter. In this work, the smart inverter is rated at $ 15 \% $ higher than the maximum active power rating of the DGs. Let, the apparent power rating of the DG connected to phase $p$ of node $i$ be $s_{DG,i}^{rated,p}$ and the forecasted active power generation be $p_{DG,i}^p$. The available reactive power $q_{DG,i}^p$ support  from the smart inverters are given by (14). Since $s_{DG,i}^{rated,p}$ and $p_{DG,i}^p$ is known, (14) is simply a box constraint. 

 \begin{equation} \label{eq14}
\small
 -\sqrt{(s_{DG,i}^{rated,p})^2 - (p_{DG,i}^p)^2} \leq q_{DG,i}^p \leq \sqrt{(s_{DG,i}^{rated,p})^2 - (p_{DG,i}^p)^2}
\end{equation}

\subsection{Voltage-Dependent Load Models}
The most widely acceptable load model is the ZIP model which is a combination of constant impedance (Z), constant current (I) and constant power (P)) characteristics of the load \cite{kers}. The mathematical representation of the ZIP model for the load connected at phase $p$ of bus $i$ is given by (15)-(16).

\begin{eqnarray} \label{eq15}
p_{L,i}^p = p_{i,0}^p \left[k_{p,1} \left(\dfrac{V_i^p}{V_0}\right)^2 + k_{p,2} \left(\dfrac{V_i^p}{V_0}\right)+ k_{p,3}\right]\\
q_{L,i}^p = q_{i,0}^p \left[k_{q,1} \left(\dfrac{V_i^p}{V_0}\right)^2 + k_{q,2} \left(\dfrac{V_i^p}{V_0}\right)+ k_{q,3}\right]
\end{eqnarray}

where, $k_{p,1}+k_{p,2}+k_{p,3} = 1$, $k_{q,1}+k_{q,2}+k_{q,3} = 1$, $p_{i,0}^p$ and $q_{i,0}^p$ are per-phase load consumption at nominal voltage $V_0$.

The ZIP load model represented in (15)-(16) are a function of both $V_i^p$ and $v_i^p = (V_i^p)^2$. Including (15) and (16) to OPF formulation will make (8) and (9), earlier linear in $v_i^p$, nonlinear. Here, we develop an equivalent load model for voltage-dependent loads using the definition of CVR factor. Next, an equivalence between ZIP parameters and proposed CVR factors is obtained.

CVR factor is defined as the ratio of percentage reduction in active or reactive power to the percentage reduction in bus voltage. Let CVR factor for active and reactive power reduction be $CVR_{p}$, and $CVR_{q}$, respectively defined in (17).

\begin{eqnarray} \label{eq17}
CVR_{p} = \dfrac{d p_{L,i}^p}{p_{i,0}^p} \dfrac{V_0}{dV_i^p} \hspace{0.2cm} \text{and} \hspace{0.2cm} CVR_{q} = \dfrac{d q_{L,i}^p}{q_{i,0}^p} \dfrac{V_0}{dV_i^p}
\end{eqnarray}

where, $p_{L,i}^p = p_{i,0}^p + d p_i^p$ and $q_{L,i}^p = q_{i,0}^p + d q_i^p$.
Furthermore, $v_i^p = (V_i^p)^2$. Therefore, $d v_i^p =  2V_i^p dV_i^p$. Assuming $V_i^p \approx V_0$ and $d v_i^p = v_i^p-(V_0)^2$, we obtain:

\begin{eqnarray} \label{eq18}
p_{L,i}^p = p_{i,0}^p + CVR_{p}\dfrac{p_{i,0}^p}{2}\left(\dfrac{v_i^p}{V_0^2}-1\right)\\
q_{L,i}^p = q_{i,0}^p + CVR_{q}\dfrac{q_{i,0}^p}{2}\left(\dfrac{v_i^p}{V_0^2}-1\right)
\end{eqnarray}

Note that the CVR based load model detailed in (18) and (19) is linear in $v_i^p$, thus can be easily included in approximate power flow equations (8) and (9).
The CVR factors, $CVR_{p}$ and $CVR_{q}$ are estimated from the ZIP coefficients of the load. On differentiating the ZIP model detailed in (15) and (16) and assuming $V_0 = 1$ p.u., we obtain:

\begin{equation} \label{eq20}
\dfrac{d p_{L,i}^p}{dV_i^p} = p_{i,0}^p \left(2 k_{p,1} V_i^p+ k_{p,2}\right) \text{,}
\dfrac{d q_{L,i}^p}{dV_i^p} = q_{i,0}^p \left(2 k_{q,1} V_i^p+k_{q,2}\right)
\end{equation}

Using (17) and (20) and assuming $V_i^p \approx V_0$, we obtain (21). Using (21), the CVR factors for customer loads can be obtained from the ZIP coefficients.

\begin{equation} \label{eq21}
CVR_p = 2 k_{p,1} + k_{p,2} \hspace{0.1cm} \text{and} \hspace{0.1cm} CVR_q = 2 k_{q,1} + k_{q,2}
\end{equation}

The proposed load model should accurately represent the ZIP load model. It is thoroughly validated in authors' prior work detailed in \cite{paper1} that the characteristics of the CVR based load model (detailed in this section) closely matches the ZIP model for the allowable ANSI voltage ranges i.e. $0.95-1.05$ pu . This proves that the proposed load model in (18)-(19), which is linear in $v_i^p$, can be used instead of the ZIP load model. 
 \begin{figure*}[t]
\centering
\includegraphics[width=6.3 in]{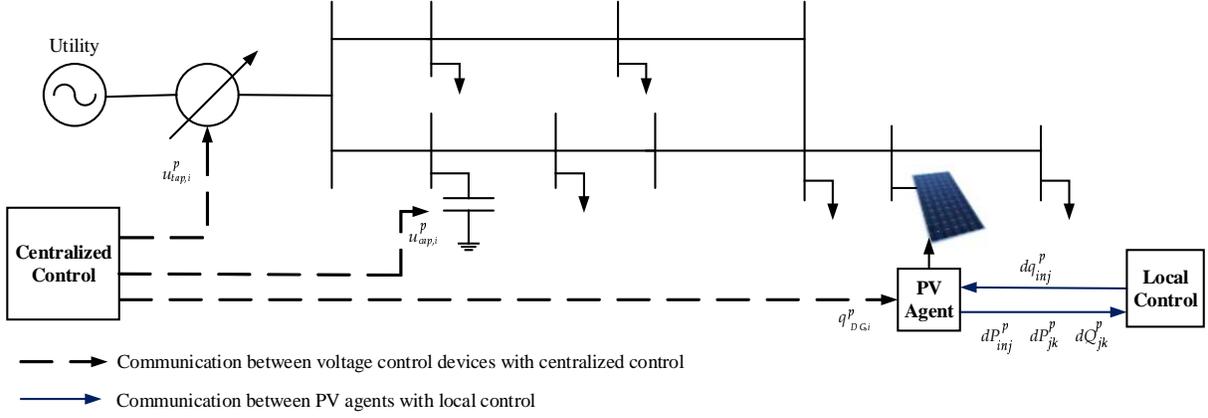}
	\caption{\footnotesize Integrating Centralized and Local Control.}
	\label{fig:2}
\end{figure*}

\section{Two-Timescale Coordinated Centralized and Local Voltage Control Framework}
The proposed two-timescale approach to coordinate the centralized and local controls is shown in Fig. 2. The objective is to achieve CVR objective by optimally controlling voltage control devices to reduce nodal voltages and consequently  help reduce feeder power consumption due to voltage-dependent loads. Note that the nodal voltages should be maintained within the recommended ANSI limits (American National Standard for Electric Power Systems and Equipment). The problem is formulated as a mixed integer linear program (MILP) where integer decision variables are introduced due to voltage regulators taps positions, and capacitor banks switches. The desired controls obtained after solving MILP are implemented to reduce feeder voltages. The centralized control decisions generate a reference voltage for each feeder node that, if maintained, will help optimally minimize the power consumption from voltage dependent loads thus help achieve the CVR objective. In order to include the effects of variations in PV generation, the centralized control is supplemented with the local control of smart inverters. The local control help to maintain the nodal voltages close to the reference voltages obtained  using the centralized controller.

\subsection{Centralized Control}
The objective of the centralized control is to minimize  the active power demand from the substation by controlling system's voltage control devices including voltage regulators, capacitor banks and smart inverters. The control of legacy devices will introduce integer variables in the optimization problem. We use the three-phase linear power flow model detailed in Section 2. This results in a MILP problem as described below. Note that the centralized control is executed at every 15-min interval. 

\begin{equation}
\small
\text{Minimize} \sum_{p \in \phi_s, j:s \rightarrow j}{P^{p}_{sj}(t)}	\end{equation}
Subject to: 
\begin{flalign}\label{eq23}
            &P_{ij}^{pp}(t) = \sum_{k:j \rightarrow k}P_{jk}^{pp}(t) + p_{L,j}^p(t)- p_{DG,i}^p(t) \hspace{0.1cm} \forall i\in\mathcal{N} \\
            &Q_{ij}^{pp}(t) = \sum_{k:j \rightarrow k}Q_{jk}^{pp}(t) + q_{L,j}^p(t)- q_{DG,i}^p(t) - q_{C,i}^p \hspace{0.1cm} \forall i\in\mathcal{N} \\
            &v_j^p(t) = v_i^p(t) - \sum_{q \in \Phi_j}{2 \mathbb{Re}\left[S_{ij}^{pq}(t)(z_{ij}^{pq})^*\right]} \hspace{0.1cm} \forall j \in \mathcal{N}  \\
            &p_{L,i}^p(t) = p_{i,0}^p(t) + CVR_{p}(t)\dfrac{p_{i,0}^p(t)}{2}(v_i^p(t)-1)  \forall i\in\mathcal{N_L} \\
            &q_{L,i}^p(t) = q_{i,0}^p(t) + CVR_{q}(t)\dfrac{q_{i,0}^p(t)}{2}(v_i^p(t)-1)  \forall i\in\mathcal{N_L} \\
            &v_j^p(t) = A_i^p(t) v_i^p(t) \hspace{0.3cm} \forall (i,j) \in \mathcal{E_T} \\
            &A_i^p(t) = \sum\limits_{i=1}^{32} B_i u_{tap,i}^p(t), \sum\limits_{i=1}^{32} u_{tap,i}^p(t) = 1 \forall (i,j) \in \mathcal{E_T}  \\
            &q_{C,i}^{p}(t) = u_{cap,i}^p(t)  q_{cap,i}^{rated,p} v_i^p(t)  \hspace{0.2cm} \forall (i) \in \mathcal{N_C} \\
            &q_{DG,i}^p(t) \leq \sqrt{(s_{DG,i}^{rated,p})^2 - (p_{DG,i}^p)^2(t)} \hspace{0.2cm} \forall (i) \in \mathcal{N_{DG}} \\
            & q_{DG,i}^p(t) \geq -\sqrt{(s_{DG,i}^{rated,p})^2 - (p_{DG,i}^p)(t)^2} \hspace{0.2cm} \forall (i) \in \mathcal{N_{DG}}\\
            &(V_{min})^2\leq v_i^p(t) \leq (V_{max})^2 \hspace{12pt} \forall i\in \mathcal{N}
 \end{flalign}

The branch capacity for a line is defined as the maximum permissible kVA capacity calculated based on its ampacity. Thus, the loading of a line should not exceed its maximum permissible capacity as defined below:
\begin{equation}
      (P_{ij}^{pp})^2 (t)  + (Q_{ij}^{pp})^2 (t) \leq ((S_{ij}^{pp})^{max})^2
\end{equation}
To avoid the resulting nonlinearity in the optimal power flow formulation, the quadratic constraints in (34) is approximated as linear constraints using a polygon based linearization method first proposed in \cite{ahmadi2014linear}.
For each line $(i,j)$ the linear constraints can be formulated as:
\begin{equation}
        \begin{array}{cc}
         & -\sqrt{3} (P_{ij}^{pp} (t) +  R_{ij}^{pp} ) \leq  Q_{ij}^{pp} (t) \leq -\sqrt{3} (P_{ij}^{pp} (t) -  R_{ij}^{pp} )  \\
         & \frac{-\sqrt{3}}{2}   R_{ij}^{pp}  \leq  Q_{ij}^{pp} (t) \leq \frac{\sqrt{3}}{2}   R_{ij}^{pp}  \\
         &  \sqrt{3} (P_{ij}^{pp} (t) - R_{ij}^{pp} ) \leq  Q_{ij}^{pp} (t) \leq \sqrt{3} (P_{ij}^{pp} (t) +  R_{ij}^{pp} )  \\
        \end{array}
   \end{equation}
  where, the radius of a hexagon $R_{ij}^{pp}$ is obtained using 
         \begin{equation}
             R_{ij}^{pp} = (S_{ij}^{pp})^{max} \sqrt{\frac{2 \pi/6}{sin(2 \pi/6)}}  \hspace{1 cm}  (i,j) \in \mathcal{E}
         \end{equation}

The problem objective (22) is to minimize the summation of active power demand from the substation in all the three phases. Constraints (23) - (25) are the linear AC power flow constraints. Constraints (26)-(27) detail nodal load demand based on the load model detailed in Section 3.4. Constraints (28)-(29) represent model for voltage regulator control, and (30) represents capacitor model. Constraints (31)-(32) are the limits on the reactive power support provided by the PV smart inverters. Constraint (33) represents the ANSI voltage limits at each node. Constraint (35)-(36) represents the linear branch capacity limits.

\subsection{Local Smart Inverter Control}
The objective of the local control is to reduce the voltage deviations at the nodes wrt. the reference nodal voltages, $V_{i,ref}^p$, obtained using the centralized controller. The objective for the local control is as follows:
 \begin{equation} \label{eq34}
\text{Minimize} \hspace{1 cm} {|V_i^{p}(t)- V_{i,ref}^p(t)|}	\hspace{1 cm} \forall i\in\mathcal{N_{DG}} 
\end{equation}
where, $V_i^{p}(t)$ is the actual nodal voltage at phase $p$ of node $i$. 
 
In  section 5, two local control methods are proposed to achieve the local control objective. The local control updates the smart inverter's reactive power support to mitigate the effects of local PV generation variability on nodal voltage fluctuation. Note that only local measurements available at the point of connection for the PV smart inverter are used to obtained the control decisions. 

\subsection{Infrastructure to Implement the Proposed Framework}
The required infrastructure to implement the proposed framework is detailed in this section. The centralized control is located at the utility control center that receives measurement information using an advanced distribution management system (ADMS) also located at the utility control center. ADMS includes: (1) distribution supervisory control and data acquisition (DSCADA) - to gather information on network device statuses (capacitor banks, voltage regulator, switches), (2)  advanced meter infrastructure (AMI) - to obtain load data, (3) distributed energy resource management system (DERMS) or distributed energy resources (DERs) aggregators - to obtain the information on DERs (generation and available reactive power). For behind-the-meter resources (such as roof-top PVs), we assume that an aggregator is available to communicate and control the individual behind-the-meter smart inverters. The centralized controller after receiving the information on loads, utility-scale/community solar, aggregated PVs, performs optimization with the objective of minimizing the power demand from the substation. It solves for the optimal capacitor bank switch statuses, voltage regulator tap positions, and smart inverter reactive power support. The control signals are dispatched to capacitor banks switch, voltage regulators, and DERs \cite{mani}. For aggregated PVs, the decision from the central controller is distributed among the PVs and communicated to the individual smart inverters by DER aggregators. It is assumed that the PVs and smart inverters are controllable and have the communication and measuring capabilities at the point of connection. Mostly, the load demand measurements are gathered every 15-min time interval using AMI. Therefore, we assume that the centralized controller operates every 15-min interval. 

The centralized control decisions are communicated to individual grid voltage control devices at every 15-min interval. Thus, the centralized control signals are fixed for the specified 15-min time-window. As mentioned before, the variability in PV power generation will lead to voltage fluctuations within the 15-min decision interval specified for the centralized controller. A fast local controller is implemented at each smart inverter to update their reactive power support in response to local changes in PV generation. The proposed local controller operates in sub-intervals (from real-time to every 1-min interval depending upon the granularity of the local measurements) and are designed to reduce the deviations between the actual nodal voltages and the reference voltages specified by the centralized controller for the given 15-min interval.

\section{Proposed Methods for Local Control}   
The intermittent nature of PV generators causes voltage fluctuation in the distribution system. The voltage reference for a node obtained from the centralized controller must be maintained to realize the CVR objective. In this section, the factors which causes  the voltage fluctuations in the distribution systems are discussed. Later, two local voltage control methods are proposed to reduce the nodal voltage deviations. 

\subsection{PV Variability}
PVs have lately emerged as the most prominent DGs connected at the distribution level. At the distribution level, both small-scale (roof-top PVs) and utility-scale PV generation plants are being widely deployed throughout the nation. The PV power generation is, in general, variable due to the variations in received solar irradiance. Typically, for a clear sky day, it is assumed that the PV generation follows a smooth and nearly parabolic profile (see Fig. 3). However, during a cloudy day, the PV irradiance can rapidly fluctuate leading to a rapidly varying PV generation profile as shown in Fig. 3. Due to intermittent nature of PV power production, there can be a sudden increase or decrease in nodal voltages causing voltage fluctuations in the distribution system. The voltage fluctuations can lead to voltage quality problems, and/or lead to an increased wear and tear of system's legacy voltage devices such as capacitor banks and voltage regulators \cite{dubey2017impacts}.  Therefore, it is of interest to minimize the nodal voltage fluctuations resulting from intermittent PV generation. 

In the proposed work, the reactive power support from the smart inverters is utilized to reduce voltage fluctuations in the distribution system due to random variations in PV power generation. These are referred to as local control methods for smart inverter operation. Furthermore, it is ensured that the proposed local control approach is coordinated with the centralized voltage control approach as detailed in Section 4.1. Note that we have used three-sigma rule to generate the variable PV power output \cite{threesigma}. The normalized PV power production with 30 \%  and 70 \% variability is shown in Fig. 3. 

\begin{figure}[t]
\centering
\includegraphics[width=3.2 in]{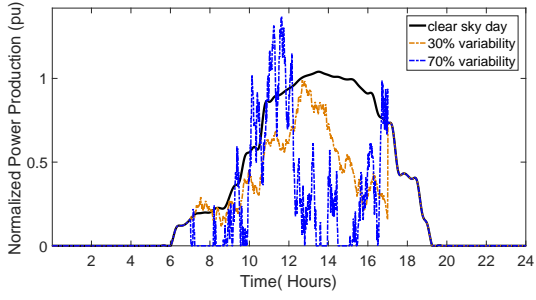}
 \vspace{-0.3cm}
	\caption{\footnotesize PV generation profile with variability.}
\vspace{-0.5cm}
	\label{fig:3}
\end{figure}

\subsection{Local Smart Inverter Control - using Equivalent Thevenin Impedance Method}
The problem objective is to minimize the voltage deviation at distribution system nodes caused due to the change in active power generation by PVs. We assume PV system as a current source as shown in Fig. 4. The voltage at a node $i$ for phase $p$, $V^p_{i}$, is given by (\ref{eq35}) where, $I^p_{inj}$ is the current injected by the PV system and $Z_{i}$ is the equivalent Thevenin impedance (positive sequence impedance) at bus $i$.  

\begin{equation}\label{eq35}
    V^p_{i} = Z_{i}I^p_{inj} = (R_{i}+\iota X_{i})I^p_{inj} \hspace{0.5 cm} p\in \{a,b,c\}
\end{equation}

The current injected at the node $i$, $I^p_{inj}$, is obtained using (\ref{eq36}), where $P^p_{inj}$ and $Q^p_{inj}$ are the per-phase active and reactive power injected at node $i$ at nominal voltage level, $V^p_{nom}$. 

\begin{equation}\label{eq36}
    I^p_{inj} = \frac{(P^p_{inj}+\iota Q^p_{inj})^*}{V^p_{nom}} \hspace{0.5 cm} p\in \{a,b,c\}
\end{equation}

Next, we obtain the expression for voltage deviation at a node as a function of change in the active and reactive power injected at the node in (\ref{eq40}). The deviation in nodal voltage due to the change in active and reactive power is calculated using (41)-(42).
\begin{eqnarray} \label{eq40}
dV^p_i = \frac{\partial V^p_i}{\partial P^p_{inj}}dP^p_{inj} + \frac{\partial V^p_i}{\partial Q^p_{inj}}dQ^p_{inj}  \hspace{0.4 cm} p\in \{a,b,c\}\\
\frac{\partial V^p_i}{\partial P^p_{inj}} = \frac{(R_{i}+ \iota X_{i})}{V^p_{nom}} \hspace{1.7 cm} p\in \{a,b,c\}\\
\frac{\partial V^p_i}{\partial Q^p_{inj}} = -\iota \frac{(R_{i}+ \iota X_{i})}{V^p_{nom}} \hspace{1.5 cm} p\in \{a,b,c\}
\end{eqnarray}

\begin{figure}[t]
\centering
\includegraphics[width=3.3 in]{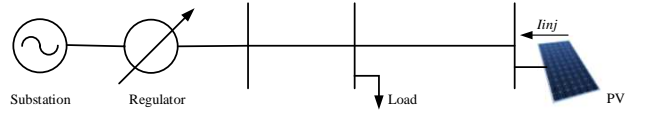}
 \vspace{-0.2cm}
	\caption{\footnotesize Distribution system connected to PV}
\vspace{-0.3cm}
	\label{fig:4}
\end{figure}

Using (40)-(42), we obtain the expression for the voltage deviation as a function of the equivalent Thevenin impedance and the change in active and reactive power injected at the node in (\ref{eq41}). 
\vspace{-0.2cm}
\begin{equation} \label{eq41}
\begin{split}
    dV^p_i = \left(\frac{R_{i}}{V^p_{nom}}dP^p_{inj}+ \frac{X_{i}}{V^p_{nom}}dQ^p_{inj}\right) \\
    + \iota \left(\frac{X_{i}}{V^p_{nom}}dP^p_{inj}- \frac{R_{i}}{V^p_{nom}}dQ^p_{inj}\right)
\end{split}
\end{equation}

\begin{figure}[t]
\centering
\includegraphics[width=3.2 in]{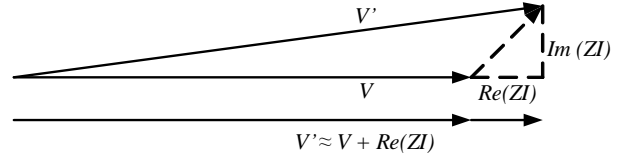}
 \vspace{-0.2cm}
	\caption{\footnotesize Phasor diagram for change in voltage}
\vspace{-0.4cm}
	\label{fig:5}
\end{figure}

The change in voltage at a node due to the change in current injected at the node is the vector sum of nodal voltage, $V$, and impedance times injected current, $ZI$. Since the angle difference between the source node and load node voltages is small, as shown in Fig. 5, the change in the nodal voltage is approximately equal to the real part of the product of impedance and current \cite{kers}. Hence, the imaginary part of equation (\ref{eq41}) is ignored. 

Next, in order to maintain the nodal voltage constant and equal to the value obtained using the centralized control, the change in voltage $dV^p_i$ should be made equal to zero. Therefore, equating  real part of (\ref{eq41}) to zero, the desired reactive power support required to maintain nodal voltage constant is obtained in (\ref{eq:44}). From (\ref{eq:44}),  it can be observed that the change in reactive power required to maintain the voltage at a node is directly proportional to the change in active power injected at the node scaled by $R/X$ ratio at the node. 
\begin{equation} \label{eq:44}
    dq^p_{inj}(t) = -\frac{R_{i}}{X_{i}}dP^p_{inj}(t)
\end{equation}

The local control law stated in (\ref{eq:44}) requires the Thevenin impedance as seen from the node and the change in active power generation from PV wrt. the forecasted value (that was used when solving centralized control problem).The Thevenin impedance is the positive sequence impedance for each bus in the distribution system. We use OpenDSS to obtain the positive sequence impedance at each bus of the given distribution system. Please refer to \cite{kers} for a detailed explanation on the approach used by OpenDSS to calculate the positive sequence impedance. The change in active power injected by PVs are measured at each node. Using the positive sequence impedance and change in active power, the calculated value of $dq^p_{inj}$ is used to update the reactive power support from smart inverter that was previously calculated using the centralized controller, $q^p_{inj}$. The smart inverter reactive power support at every interval of local control is obtained using $   q^p_{inj,new} = q^p_{inj}+dq^p_{inj}$. 

The net reduction in nodal voltage deviations wrt. the voltage reference is dependent on the Thevenin impedance at the PV location. Here, the Thevenin impedance is calculated after ignoring the effects of shunt devices in the distribution system. Hence, the calculated values are not the exact values for all load and generation conditions. Also, in this method, the compensation provided by the smart inverters connected at other nodes is ignored. Therefore, there is no coordination among the smart inverters when trying to reduce the nodal voltage fluctuations. The local control decisions, therefore, are approximate.

\subsection{Local Smart Inverter Control - using Power Flow Measurements}
In this section, a measurement-based local smart inverter method is proposed to mitigate the voltage deviations caused by variations in PV power generation. The proposed method is based on the linear three-phase power flow equations described in Section 2.2. It should be noted that the proposed method includes the effects of local control actions from other smart inverters and results in a better minimization of nodal voltage deviations.

The deviation in nodal voltages due to variation in PV generation can be obtained by differentiating the three-phase voltage equations detailed in (10). The differentiated voltage equation is shown in (\ref{eq42}), where, $\theta^{pq}$ is the angle difference between phase voltages which is assumed to be a constant value, see (5). 
\vspace{-0.1cm}
\begin{equation} \label{eq42}
\small
\begin{split}
    dv_i^p - dv_j^p = \sum_{q \in \phi_ j}  2 \left(r_{ij}^{pq}dP_{ij}^{pq}cos(\theta^{pq}) + x_{ij}^{pq}dQ_{ij}^{pq}sin(\theta^{pq}) \right) \end{split}
\end{equation}

Next, same as in the previous section, we equate the changes in nodal voltages to zero i.e., $dv_i^p = dv_j^p = 0$. We obtain the mathematical relation between the change in reactive power flow as a function of change in active power flow, detailed in (\ref{eq43}). 

\vspace{-0.1cm}
\begin{equation}  \label{eq43}
\small
\begin{bmatrix}
r_{ij}^{aa} & A_1 & A_3 \\
B_1  & r_{ij}^{bb} & B_3 \\
C_1  & C_3 & r_{ij}^{cc}  \\
\end{bmatrix}
\begin{bmatrix}
dP_{ij}^{a} \\
dP_{ij}^{b} \\
dP_{ij}^{c} \\
\end{bmatrix} + 
\begin{bmatrix}
x_{ij}^{aa} & A_2 & A_4 \\
B_2  & x_{ij}^{bb} & B_4 \\
C_2  & C_4 & x_{ij}^{cc}  \\
\end{bmatrix}
\begin{bmatrix}
dQ_{ij}^{a} \\
dQ_{ij}^{b} \\
dQ_{ij}^{c} \\
\end{bmatrix}
=
\begin{bmatrix}
0 \\
0 \\
0 \\
\end{bmatrix}
\end{equation}

where, 
$A_1 = (-\frac{1}{2}r_{ij}^{ab} + \frac{\sqrt{3}}{2}x_{ij}^{ab}) $ , $A_3 = (-\frac{1}{2}r_{ij}^{ac} - \frac{\sqrt{3}}{2}x_{ij}^{ac}) $ \\
$A_2 = (-\frac{\sqrt{3}}{2}r_{ij}^{ab} - \frac{1}{2}x_{ij}^{ab}) $ , $A_4 = (\frac{\sqrt{3}}{2}r_{ij}^{ac} - \frac{1}{2}x_{ij}^{ac}) $  \\
$B_1 = (-\frac{1}{2}r_{ij}^{ab} - \frac{\sqrt{3}}{2}x_{ij}^{ab}) $ , $B_3 = (-\frac{1}{2}r_{ij}^{bc} + \frac{\sqrt{3}}{2}x_{ij}^{bc}) $ \\
$B_2 = (\frac{\sqrt{3}}{2}r_{ij}^{ab} - \frac{1}{2}x_{ij}^{ab}) $  , $B_4 = (-\frac{\sqrt{3}}{2}r_{ij}^{bc} - \frac{1}{2}x_{ij}^{bc}) $ \\
$C_1 = (-\frac{1}{2}r_{ij}^{ac} + \frac{\sqrt{3}}{2}x_{ij}^{ac}) $ , $C_3 = (-\frac{1}{2}r_{ij}^{bc} - \frac{\sqrt{3}}{2}x_{ij}^{bc}) $ \\
$C_2 = (-\frac{\sqrt{3}}{2}r_{ij}^{ac} - \frac{1}{2}x_{ij}^{ac}) $ , $C_4 = (\frac{\sqrt{3}}{2}r_{ij}^{bc} - \frac{1}{2}x_{ij}^{bc}) $ \\

Due to the change in active and reactive power generation by the PVs there will be change in the branch/line power flow. The relationship for the change in power flow in the line due to change in active and reactive power injected by the PVs is obtained by differentiating (\ref{eq:8}) and (9). The change in active and reactive power flow in the lines for each phase are shown in (\ref{eq44}) and (\ref{eq45}), respectively. 

\begin{equation}  \label{eq44}
\begin{bmatrix}
dP_{ij}^{a}  \\
dP_{ij}^{b} \\
dP_{ij}^{c}  \\
\end{bmatrix} -
\begin{bmatrix}
dp_j^{a} \\
dp_j^{b}  \\
dp_j^{c}  \\
\end{bmatrix}
=
\begin{bmatrix}
\sum_{k:j \rightarrow k} dP^{a}_{jk} \\
\sum_{k:j \rightarrow k} dP^{b}_{jk} \\
\sum_{k:j \rightarrow k} dP^{c}_{jk} \\
\end{bmatrix}
\end{equation}
 \vspace{-0.2cm}
\begin{equation}  \label{eq45}
\begin{bmatrix}
dQ_{ij}^{a}  \\
dQ_{ij}^{b} \\
dQ_{ij}^{c}  \\
\end{bmatrix} -
\begin{bmatrix}
dq_j^{a} \\
dq_j^{b}  \\
dq_j^{c}  \\
\end{bmatrix}
=
\begin{bmatrix}
\sum_{k:j \rightarrow k} dQ^{a}_{jk} \\
\sum_{k:j \rightarrow k} dQ^{b}_{jk} \\
\sum_{k:j \rightarrow k} dQ^{c}_{jk} \\
\end{bmatrix}
\end{equation}

Next, in order to reduce the nodal voltage deviations, we use (\ref{eq43})-(\ref{eq45}) to obtain (\ref{eq46}) that provides the required reactive power support from each smart inverter.

\vspace{-0.3cm} 
\begin{equation}\label{eq46}
\begin{split}
\begin{bmatrix}
dq_j^{a} \\
dq_j^{b}  \\
dq_j^{c}  \\
\end{bmatrix}
= \left(- [A_{ij}]^{-1} + \sum_{k:j \rightarrow k}[A_{jk}]^{-1} \right)
\begin{bmatrix}
\sum_{k:j \rightarrow k} dP^{a}_{jk} \\
\sum_{k:j \rightarrow k} dP^{b}_{jk} \\
\sum_{k:j \rightarrow k} dP^{c}_{jk} \\
\end{bmatrix} \\
- [A_{ij}]^{-1}
\begin{bmatrix}
dp_j^{a} \\
dp_j^{b}  \\
dp_j^{c}  \\
\end{bmatrix} \\
\end{split}
\end{equation}
\vspace{-1.1cm} 
where,  

\begin{equation}
\begin{bmatrix}
A_{ij}
\end{bmatrix}  
=
\begin{bmatrix}
x_{ij}^{aa} & A_2 & A_4 \\
B_2  & x_{ij}^{bb} & B_4 \\
C_2  & C_4 & x_{ij}^{cc}  \\
\end{bmatrix}^{-1}
\begin{bmatrix}
r_{ij}^{aa} & A_1 & A_3 \\
B_1  & r_{ij}^{bb} & B_3 \\
C_1  & C_3 & r_{ij}^{cc}  \\
\end{bmatrix}
\end{equation}

Note that the inverters located downstream from a PV node are also providing the reactive power support and, therefore, affect the power flow along the line downstream from the node. Therefore, by including downstream line power flow measurement when calculating the required reactive power support for smart inverters in (\ref{eq46}), the effects of reactive power support provided by other smart inverters in the system are taken into account. Thus, (\ref{eq46}) provides a better local control to regulate nodal voltages as compared to Thevenin impedance-based method introduced in the previous section. To implement this control approach, the local smart inverter requires two local measurements: (1) local PV generation, and (2) power flow (P, Q) into the children nodes from the PV nodes.

\begin{figure}[t]
\centering
\includegraphics[width=3.3 in]{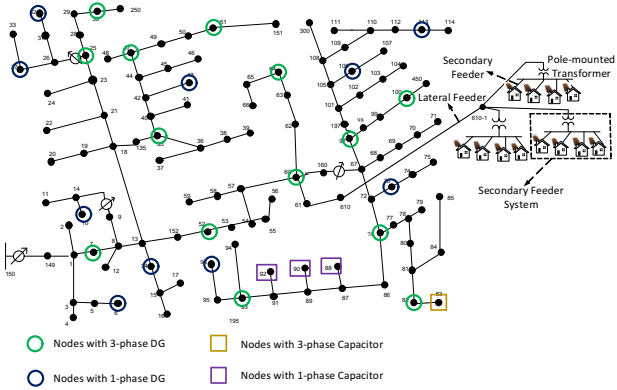}
\vspace{-0.4cm}
	\caption{\footnotesize Modified IEEE 123-bus distribution test feeder.}
\vspace{-0.7cm}
	\label{fig:6}
\end{figure}

\begin{figure*}[t]
\centering
\includegraphics[width=6.3 in ]{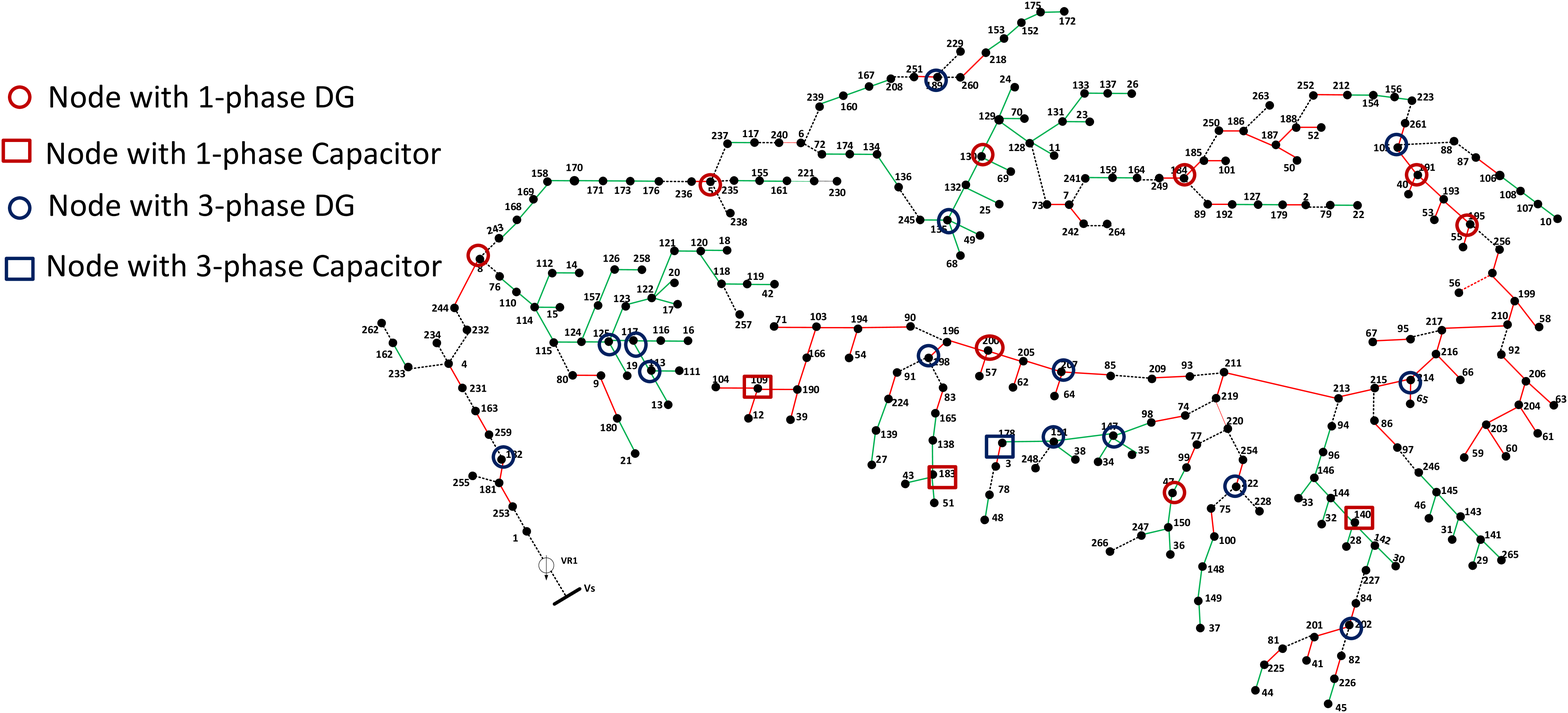}
\vspace{-0.5cm}
	\caption{Modified  R3-12.47-2 distribution test feeder.}
	\vspace{-0.5cm}
	\label{fig:7}
\end{figure*}

\section{Results and Discussions}
The proposed control framework is validated using the modified IEEE-123 bus system  and the  modified  R3-12.47-2  feeder  with simulation done on Matlab and OpenDSS. The interface between Matlab and OpenDSS is established using COM interface. For the centralized control, the problem is modeled as a MILP which is solved using CPLEX 12.7. The local control algorithm is executed in Matlab and the obtained control variables are provided to the distribution model simulated in OpenDSS. Therefore, although control decisions are based on linearized power flow model, OpenDSS simulates actual nonlinear three-phase system and provides a suitable environment for evaluating the performance of the proposed methods on real-world distribution systems.  All the simulations are done on i7 3.41 GHz processor with 16 GB of RAM. The proposed models are scalable and suitable for real-time operation. On an average it takes less than 30 sec. to solve the MILP problem for the centralized controller for both IEEE 123-bus (with a total of 267 single-phase nodes) and the  modified  R3-12.47-2  feeder  (with a total of 860 single-phase nodes) test feeders. 

The IEEE-123 bus feeder (with a total of 267 single-phase nodes) is a medium size feeder  with several single phase lines and the single phase loads. There are four capacitor banks and four voltage regulators deployed along the feeder as shown in Fig. 6. The feeder is modified to include a total of 55 nodes populated with PVs spread across the feeder; three-phase PVs of 172.5 kVA or 69 kVA ratings and single phase PVs of 23 kVA or 11.5 kVA ratings. The  modified  R3-12.47-2  feeder (a total of 860 single-phase nodes) is a large feeder with single phase and three phase lines and loads. The feeder includes one substation voltage regulator, one three-phase 600-kVAr capacitor bank and three 100 kVAr single-phase capacitor banks. In order to implement the proposed control algorithm, the feeder is modified to include a total of 50 nodes populated with PVs spread across the feeder (see Fig. 7); three-phase PVs of ratings 690 kVA, 345 kVA, 69 kVA, 172.5 kVA, 138 kVA or 34.5 kVA and single phase PVs of ratings 23 kVA or 11.5 kVA. The customer loads are assumed to have  a CVR factor of 0.6 for active power and 3 for reactive power \cite{EPRI}. The generation profile for the PV during a clear sky day is used for MILP problem modeled in centralized control problem. The clear day sky profile should be thought of as PV forecast and is based on example PV profiles provided in OpenDSS (see Fig.3). The actual PV profiles to be used in local control include PV variability that is added to the clear sky PV profile to generate multiple PV variability cases. 

\subsection{Verification of Approximate Power Flow Formulation}
This section validates the three-phase linear AC power flow model detailed in Section 2.2. The results obtained from the proposed linear power flow model are compared with the nonlinear power flow solutions obtained using OpenDSS. Table 1 shows the largest errors in the power flow and bus voltages for the two test feeders at different loading conditions. It should be noted that although the error in the power flow obtained using linear power model are high, nodal voltages are well approximated. The maximum errors in nodal voltages for the  IEEE-123 bus and the  modified  R3-12.47-2  feeder  at peak load condition are 0.007 pu and 0.002 pu, respectively. The approximated and actual nodal voltages for 123-bus system are also shown using voltage profile plots in Fig. 8. Note that approximated voltage closely match the actual nodal voltages for each phase. Since optimizing for CVR requires a power flow model that accurately represents nodal voltages so that the voltage can be driven towards the lower limits specified by ANSI std. (American National Standard for Electric Power Systems and Equipment) . Since linear power flow model approximates voltages well, it can be used to model optimization problem to meet the CVR objective. 

\begin{figure}[t]
\centering
\includegraphics[width=3.3 in]{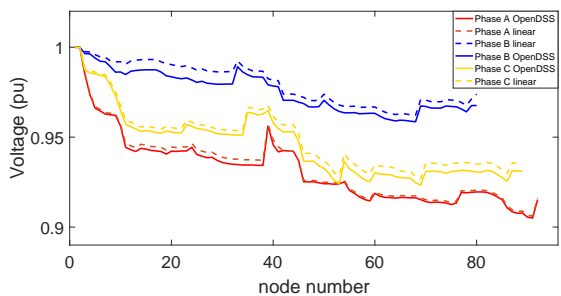}
	\caption{\footnotesize IEEE 123-bus: comparing nodal voltages obtained using linearized power flow model and OpenDSS at peak load condition}
	\label{fig:8}
	\vspace{-0.6cm}
\end{figure} 

\begin{table}[t]
\centering
\caption{Largest error in linear power flow wrt OpenDSS solutions}
\vspace{0.1cm}
\begin{tabular}{c|c|c|c}
\hline
\hline
Test feeder  & Loading($\%$)  & Error in $S_{flow}$($\%$) & Error in V(pu.)\\
\hline
IEEE 123 Bus & 50 $\%$ & 2.76 &  0.001 \\
\hline
IEEE 123 Bus & 75 $\%$ & 4.25 &  0.004 \\
\hline
IEEE 123 Bus & 100\% & 5.76 &  0.007 \\
\hline
R3-12.47-2  & 50 $\%$ & 1.1 &  0.0001 \\
\hline
R3-12.47-2  & 75 $\%$ & 2.42 &  0.001 \\
\hline
R3-12.47-2  & 100 $\%$ & 3.68 &  0.002 \\
\hline
\hline
\end{tabular}
\vspace{-0.4cm}
\end{table}

\subsection{Effects of PV Variability on Legacy Device Operations}
While implementing the centralized control, the PV generation is assumed to be constant during the 15-min time interval. However, it is well-known that PV generation is variable in smaller time intervals. To generate variable PV profiles, we have used the PV profile for a clear sky day as shown in Fig. 3. The variable generation is obtained by adding random numbers to the PV profile for the clear sky day (see Fig. 3). The variability is quantified using the three-sigma rule.

In this section, we analyze the effects of variable PV generation on the operation of legacy voltage control devices operating in autonomous control mode. Note that the proposed local control methods for smart inverter are not implemented. The simulation is performed using IEEE 123-bus system for every 1-min time interval. The PVs are operating at unity power factor and the capacitor banks and voltage regulators are used for the voltage regulation. Table 2 shows the number of switching operations in a day for the four capacitor banks and five voltage regulators installed in the test feeder. It can observed from  Table 2 that the increase in PV variability causes an increase in the number of switching operations for the legacy voltage control devices. The increased switching operations will deteriorate the performance of the legacy devices. This calls for fast control actions to mitigate the effects of PV variability on nodal voltage fluctuations. The smart inverters are a feasible option as they can respond faster than the legacy devices. Note that the centralized control is not a conducive option for the fast control of smart inverters in response to rapid changes in PV generation. The centralized control requires establishing fast communication channels between control center and smart inverters. The high cost of establishing fast communication channels, and communication delays or packet drops makes the approach infeasible for fast control of smart inverters. This makes a strong case for the local smart inverter control in order to reduce the voltage fluctuations due to rapidly varying PV generation.

 We also compare the number of legacy device operations after implementing the proposed coordinated centralized and local controller. It should be noted the number of operations for legacy device are not affected as the PV variability increases. This is because the proposed controller is programmed to mitigate the impacts of PV variability using the local control of smart inverters instead of legacy devices. Recall that the central controller obtains the optimal control statuses for the capacitor banks switches and voltage regulators taps at every 15-min interval. These controls are then frozen for the 15-min time interval and are recalculated only at the next sampling time interval. Within the 15-min interval, as the PV generation changes, the local controller operates smart inverter to compensate the effect of PV generation variability on nodal voltage variations. Thus, when the local control is implemented to mitigate the PV variability effects, the number of operation of the legacy devices do not change as these are operated using centralized controller that executes at every 15-min interval.

\begin{table}[t]
\small
        \centering
             \caption{Effect of Local Control on Switching Operation of Legacy Device in a Day}
		\label{singletable}
		\begin{tabular}{c|c|c|c|c|c}
		\hline
		\hline
		\multirow{3}{*}{*}  &  \multirow {3}{*}{0\% variability} & \multicolumn{2}{c}{30\% variability} & \multicolumn{2}{c}{70\% variability} \\ 
	    Legacy  &  &  without & with  &   without & with  \\ 
        device  &  & control  &   control &  control &   control \\
            \hline
			cap 1  & 15 & 113 & 15 & 95 & 15\\
			\hline
			cap 2  & 6 & 134 & 6 & 148 & 6 \\
            \hline
            cap 3  & 0 & 68 & 0 & 116 & 0 \\
			\hline
			cap 4  & 23 & 226 & 23 & 218 & 23 \\
			\hline
			VR 1  & 0 & 87 & 0 & 87 & 0 \\
			\hline
			VR 2  & 23 & 315 & 23 & 292 & 23 \\
			\hline
			VR 3  & 47 & 282 & 47 & 263 & 47 \\
			\hline
			VR 4  & 32 & 356 & 32 & 280 & 32 \\
			\hline
			VR 5  & 30 & 315 & 30 & 290 & 30 \\
	    \hline
	    \hline
        \end{tabular}
\vspace{-0.5cm}
\end{table}

\subsection{Centralized Control to Minimize Substation Power Demand (CVR Objective)}
The proposed centralized control is validated on the IEEE 123 bus and the  modified  R3-12.47-2  feeder. The set points for legacy devices and smart inverters are obtained using the centralized control at every 15-min time-interval. The results demonstrate that the CVR objective is achieved by maintaining the voltages at the lower levels without violating the specified ANSI voltage limits.

First, the centralized control is demonstrated using the IEEE 123-bus system. The control variables for this feeder are the switch statuses for the three single-phase capacitor banks and a three-phase capacitor bank,  tap position for all the five voltage regulators, and the reactive power support from the PV smart inverters. The total optimal power consumption supplied from the substation after implementing the proposed centralized control framework is shown in Fig. 9 and compared to the case without the centralized controller. Note that when centralized control is not implemented, it is assumed that DGs operate at unity power factor, and capacitor banks and voltage regulators operate in autonomous control modes. It can be observed from Fig. 9 that there is a reduction in power supplied by the substation. The control decisions for the legacy devices and smart inverters vary as the loading condition changes. It is observed that, at the minimum loading condition, all four capacitor banks are OFF, whereas at the maximum loading, the three-phase capacitor bank and one of the single phase capacitor bank (at node 88) is ON. The  substation voltage regulator is at -13 tap at minimum load and -8 tap at maximum load condition. The other voltage regulators also operates according to the control signals obtained from the centralized controller. The reactive power supplied/absorbed by the smart inverter varies in order to adjust the nodal voltages towards the lower ANSI voltage limits. The average reduction in active power consumption obtained after implementing the centralized control for the IEEE-123 node system is 60 kW for the day. 

To show the effect of CVR on the MV+LV distribution network, we added a secondary feeder model representing the low voltage network to the  IEEE-123 bus system at node 610 ( shown in Fig. 6). The lateral feeder is connected between node 610 and 610-1. The length of the lateral feeder is taken as 500 ft. A pole-mounted transformer is added at each phase of the lateral feeder 610-1. We have simulated four houses at each pole-mounted transformer, where each house has rated load of 6 kW. For the secondary feeder we assume there is 100 \% penetration of PV, i.e, all the house has maximum PV rating of 6 kW connected through 7.5 kVA smart inverter. The data for the LV network is shown in Table 3. \\
\begin{table}[t]\label{Table3}
		\centering
		\caption{The parameters for the components in the LV network}
		\begin{tabular}{c| c}
		\hline
		\hline
		Components & Parameters\\ 
	    \hline
		Transformers & 1-$\Phi$ 75 kVA, 2.4/0.277 kV R = 1.1 \% X = 2\% \\
		\hline
		Lines & 1-$\Phi$ ,length = 50 feet,  $\frac{X}{R} = 1.015$ \\
	\hline
	\hline
\end{tabular}
\vspace{-0.6cm}
\end{table}

 With the LV network, the centralized control operates based on the aggregated PV generation at node 610 on the primary side of pole-mounted transformer. The reactive power control obtained from centralized control is  then distributed among each PV's smart inverter according their kVA ratings. Given that  behind-the-meter PVs are usually not visible to the utility company, we assume a DER aggregator at node 610 that provides the aggregated PV generation capacity to the centralized controller and also executes the control of individual smart inverters based on the control set points obtained from the centralized controller. The centralized control calculates the required reactive power from aggregated PVs (and the control set points for other voltage control devices) by solving the optimal power flow problem. The required reactive power is communicated to the aggregator that is located at the primary side of the distribution transformer. The aggregator then distributes and communicates the required reactive power support to each PV installed at the low voltage side of the distribution feeder according to their smart inverter ratings.  The results at the maximum and minimum loading condition for MV+LV distribution network is shown in Table 4 .  Here, the simulation is performed by assuming a clear sky day. For both maximum and minimum loading condition, there is reduction in power consumption when CVR based control is implemented in the distribution system.

The secondary feeder has high $\frac{R}{X}$ ratio, thus it will increase the requirement for the reactive power support from smart inverters in order to maintain the desired voltage reference. To demonstrate this case, we present a comparison of reactive power support required when the PVs are assumed to be connected at the primary feeder-level vs for distributed PV case at the secondary feeder level (See Fig.\ref{fig:10}). It can be observed that reactive power requirement to achieve same voltage regulation is higher when PVs are connected at the secondary feeder-level.

\begin{figure}[t]
\centering
\includegraphics[width=0.45 \textwidth]{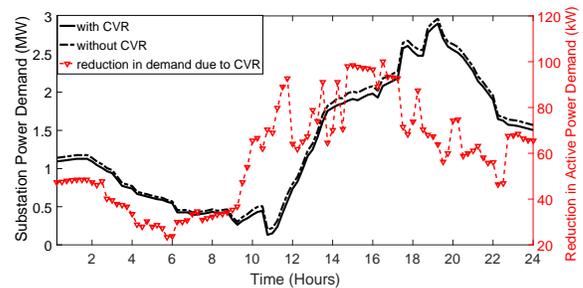}
\vspace{-0.2cm}
	\caption{\footnotesize  IEEE-123 CVR benefits observed using centralized control.}
\vspace{-0.2cm}
	\label{fig:9}
\end{figure}

\begin{figure}[t]
\centering
\includegraphics[width=0.45 \textwidth]{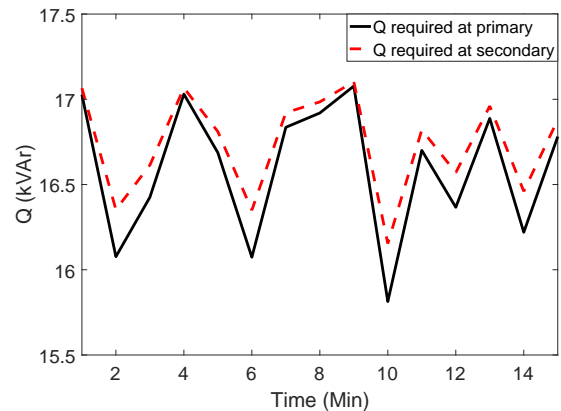}
\vspace{-0.3cm}
	\caption{\footnotesize Comparison of reactive power requirement at node 610-1.}
\vspace{-0.4cm}
\label{fig:10}
\end{figure}

\begin{table}[t]
\centering
\caption{Centralized control for IEEE-123 bus system with LV Network}
\label{singletable2}
\begin{tabular}{c|c|c}
\hline
\hline
 {Loading Condition} & {without CVR (kW)} & {with CVR (kW)} \\
  \hline
  minimum &  571.35 & 537.16 \\
  \hline
  maximum &   2158.33 & 2034.87  \\
  \hline
    \hline
\end{tabular}
\vspace{-0.5cm}
\end{table}

\begin{figure}[t]
\centering
\includegraphics[width=0.45\textwidth]{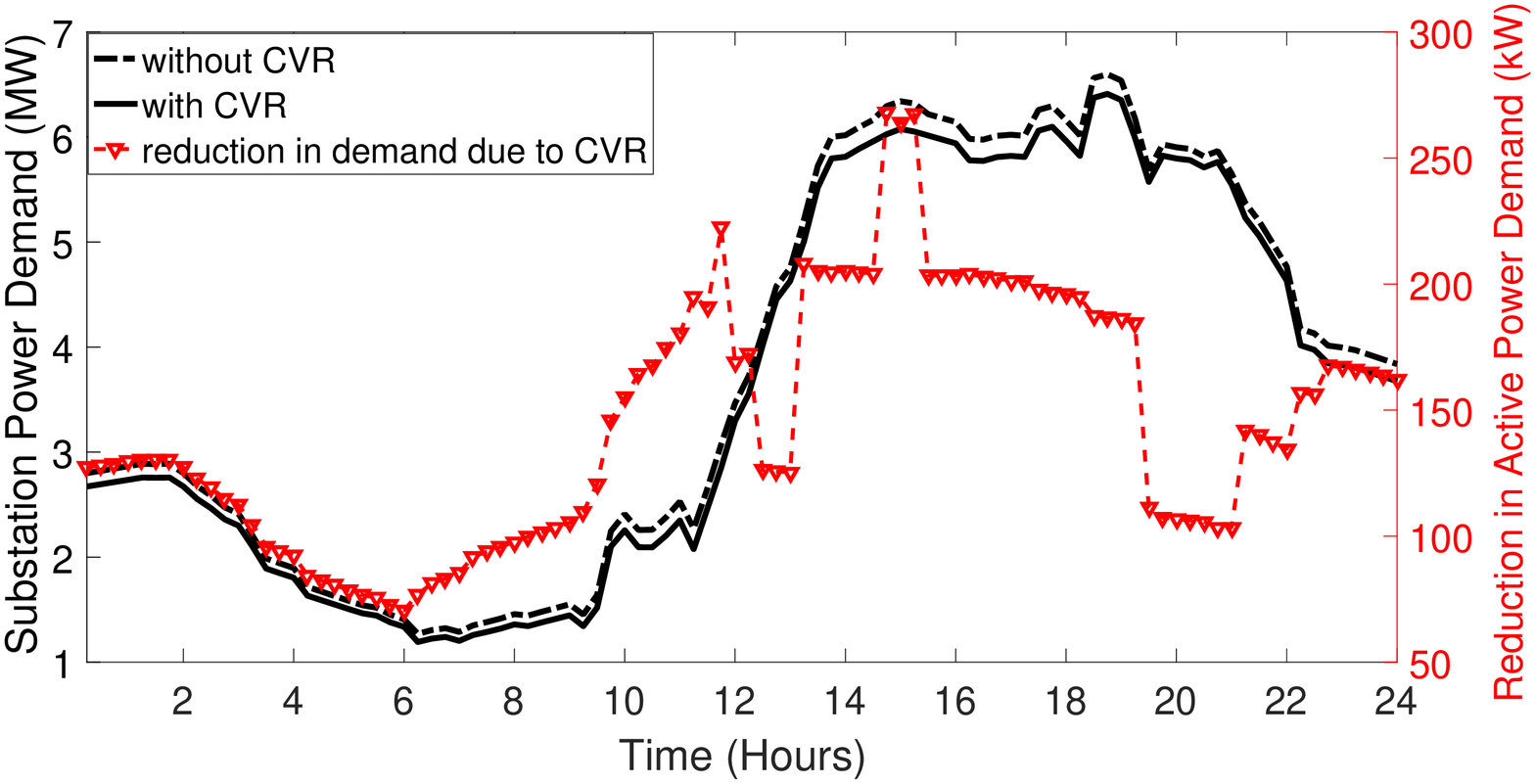}
\vspace{-0.2cm}
	\caption{\footnotesize Modified  R3-12.47-2 CVR benefits observed using  centralized control.}
\vspace{-0.5cm}
	\label{fig:11}
\end{figure}

The centralized control is also implemented for the  modified  R3-12.47-2  feeder. This test case shows the scalibility of the proposed algorithm for larger systems. At the minimum loading condition, all the capacitor banks are OFF and the substation voltage regulator is operating at -6 tap for all the three phases. At the maximum loading condition, the voltage regulator shifts to -2 tap and the three-phase capacitor banks is ON. The single-phase capacitor for phase A and C is ON, however, for phase B is OFF. The reactive power supplied/absorbed  by the smart inverters varies accordingly to maintain the voltages at the nodes towards the lower ANSI limit. The reduction in the power consumption from the substation for the  modified  R3-12.47-2 feeder is shown in Fig. \ref{fig:11}. The average power reduction for the day is approximately 145 kW for the  modified  R3-12.47-2 system. 

\subsection{Local Smart Inverter Control to Mitigate Effects of PV Variability when optimizing for CVR} 
The objective of the local control is to maintain the voltage at each node in the distribution system close to the voltages obtained by the centralized controller. In the next section, the results demonstrate the effectiveness of the local control in maintaining the desired voltages and in optimally achieving the CVR objectives while considering the variations in PV generation. Note that there is no delay in executing the local control action and the local smart inverter control is  operating in conjugation with PV generation without any delay.

\subsubsection{Reduction in Voltage Deviations}
In the following section, the effect of the proposed local control in reducing the voltage deviation is discussed. In order to quantify the effect of local control on nodal voltage fluctuations, a power quality index called system average voltage fluctuation index (SAVFI) proposed in \cite{Ding} is used. Let the nodal voltage magnitude at time \textit{t} be $V_i^t$. The voltage fluctuation $ \Delta V_i(t)$ is defined as the average of the difference in the voltage magnitude between the two time steps. 

\begin{equation}
    \Delta V_i = |V_i^{t+1}-V_i^t|
\end{equation}

The SAVFI is defined as the average of voltage fluctuation for a time interval $T$. 

\begin{equation}
  SAVFI = \frac{1}{T} \sum_{n=1}^{T} \Delta V_i(t)
\end{equation}

The SAVFI value for selected nodes of the IEEE 123-bus system is shown in Table \ref{savfi123}  for 30 \% and 70 \% PV output power variability cases. It can be observed from the table the SAVFI increases as PV variability is increased. Also, it is observed that as the distance of the nodes from the substation increases the SAVFI value increases for same levels of PV variability. In order to reduce the voltage fluctuation within the 15-min time interval, both the impedance-based and the proposed power flow measurement-based local control methods are implemented at each PV location. It can be observed from the SAVFI value (see Table \ref{savfi123} ), that both  the local control are able to reduce the voltage fluctuations at a node. Also, the proposed power flow measurement-based local control method is able to reduce the voltage fluctuations more efficiently as compared to impedance-based method. This is because, the proposed power flow measurement based method takes  the reactive power support provided by other smart inverters into account. 

\begin{table}[t]
		\centering
		\caption{SAVFI for the IEEE 123-bus system at different variability level}
		\label{savfi123}
		\begin{tabular}{c|c|c|c}
		\hline
		\hline
			Node  & Without local & Impedance-based & Power flow-based \\
			number & control & control & control \\
            \hline
            \multicolumn{4}{c}{30 \% PV power variability}\\
            \hline
			7 & 0.5734 & 0.1415 & 0.1335 \\
			\hline
			25 & 1.093 & 0.174 & 0.118 \\
			\hline
			97 & 1.647 & 0.388 & 0.273 \\
            \hline
            114 & 1.738 & 0.369 & 0.30 \\
            \hline
            \multicolumn{4}{c}{70 \% PV power variability}\\
            \hline
            7 & 1.72 & 0.433 & 0.408 \\
			\hline
			25 & 3.278 & 0.535 & 0.3712 \\
			\hline
			97 & 4.94 & 1.183 & 0.835 \\
            \hline
            114 & 5.21 & 1.126 & 0.915 \\
	\hline
	\hline
\end{tabular}
\vspace{-0.4cm}
\end{table}

\begin{table}[t]
		\centering
		\caption{SAVFI for the  modified  R3-12.47-2 system at different variability level}
		\label{savfi329}
		\begin{tabular}{c|c|c|c}
		\hline
		\hline
		Node  & Without local & Impedance-based & Power flow-based \\
		number	& control & control & control \\
            \hline
            \multicolumn{4}{c}{30 \% PV output power variability}\\
            \hline
			5 & 0.7132 & 0.1738 & 0.0295 \\
			\hline
			98 & 3.255 & 0.559 &0.0865\\
			\hline
			105 & 3.145 & 0.559 & 0.103\\
            \hline
            249 & 2.987 & 1.43 &1.12 \\
            \hline
            \multicolumn{4}{c}{70 \% PV output power variability}\\
            \hline
            5 & 2.65 & 0.549 & 0.068 \\
            \hline
            98 & 11.92 & 2.18 & 0.479 \\
            \hline
            105 & 11.52 & 2.18 & 0.53\\
            \hline
            249 & 10.93 & 5.31 & 4.19\\
	\hline
	\hline
\end{tabular}
\vspace{-0.4cm}
\end{table}

Similarly, the proposed local control methods are demonstrated to reduce voltage fluctuation in the  modified  R3-12.47-2 in Table \ref{savfi329}. From Table  \ref{savfi329}, it can be observed that the SAVFI is higher for the nodes away from the substation. It can also be verified from the table that the proposed local control based on power flow measurements is relatively more effective in reducing the voltage fluctuations for different levels of PV generation variability. 

Next, the effect of local control in meeting CVR objective while avoiding any nodal voltage violations regardless of PV variability is examined (see Table \ref{effectmaxload}). First, we observe the total substation power flow obtained after implementing the centralized control decisions (at maximum load condition) for the two cases of PV variability when local control is not implemented. We also identify the cases of voltage violations resulting in the distribution feeder due to PV variability when local control is not implemented. For the IEEE-123 node system at the maximum loading condition and at 30\% power variability, three nodes in the distribution system are observed to have voltage violation if  no local control is implemented. However, after implementing local control, none of the nodes in the system observe voltage violations. Also, the substation power demand is reduced. Similarly, for 70\% PV variability case, the total number of nodes with voltage violations increases to 13. The local control ensures no voltage violations and a reduction in substation power demand. At peak-loading condition, for the  modified  R3-12.47-2 system, the number of nodes which observe voltage violation at 30\% power variability without local control is 191 whereas at 70\% variability, the number of nodes with voltage violations increases to 445. After incorporating local control for the smart inverters, none of the nodes in the system observe voltage violations for the two cases. For 70\% PV variability, there is an increase in substation power demand. This is due to additional power requirement for the feeder to supply for losses in order to maintain the feeder voltages within the required ANSI limits.  

\begin{table}[t]
		\centering
			\caption{Effect of local control at maximum loading condition}
		\label{effectmaxload}
		\begin{tabular}{c|c|c|c|c}
		\hline
		\hline
			\multirow{3}{5.0 em}{PV  Variability }  &  \multicolumn{2}{c|}{Without local control} & \multicolumn{2}{c}{With local control}\\ 
			  &  Substation & Voltage &  Substation  & Voltage \\ 
            & Power (kW)  &  Violations & Power (kW) &  Violations \\
            \hline
            \multicolumn{5}{c}{IEEE-123 bus system}\\
            \hline
			30 \% & 2216.4 & 3 & 2209.1 & 0\\
			\hline
			70 \% & 2448.6 & 13 & 2440 & 0\\
            \hline
            \multicolumn{5}{c}{Modified  R3-12.47-2 system}\\
            \hline
            30 \% & 6619.4 & 191 & 6613.3 & 0\\
			\hline
			70 \% & 6978.1 & 445 & 6985.2 & 0\\
	\hline
	\hline
\end{tabular}
\vspace{-0.3 cm}
\end{table}

\begin{figure}[t]
\centering
\includegraphics[width=0.45\textwidth]{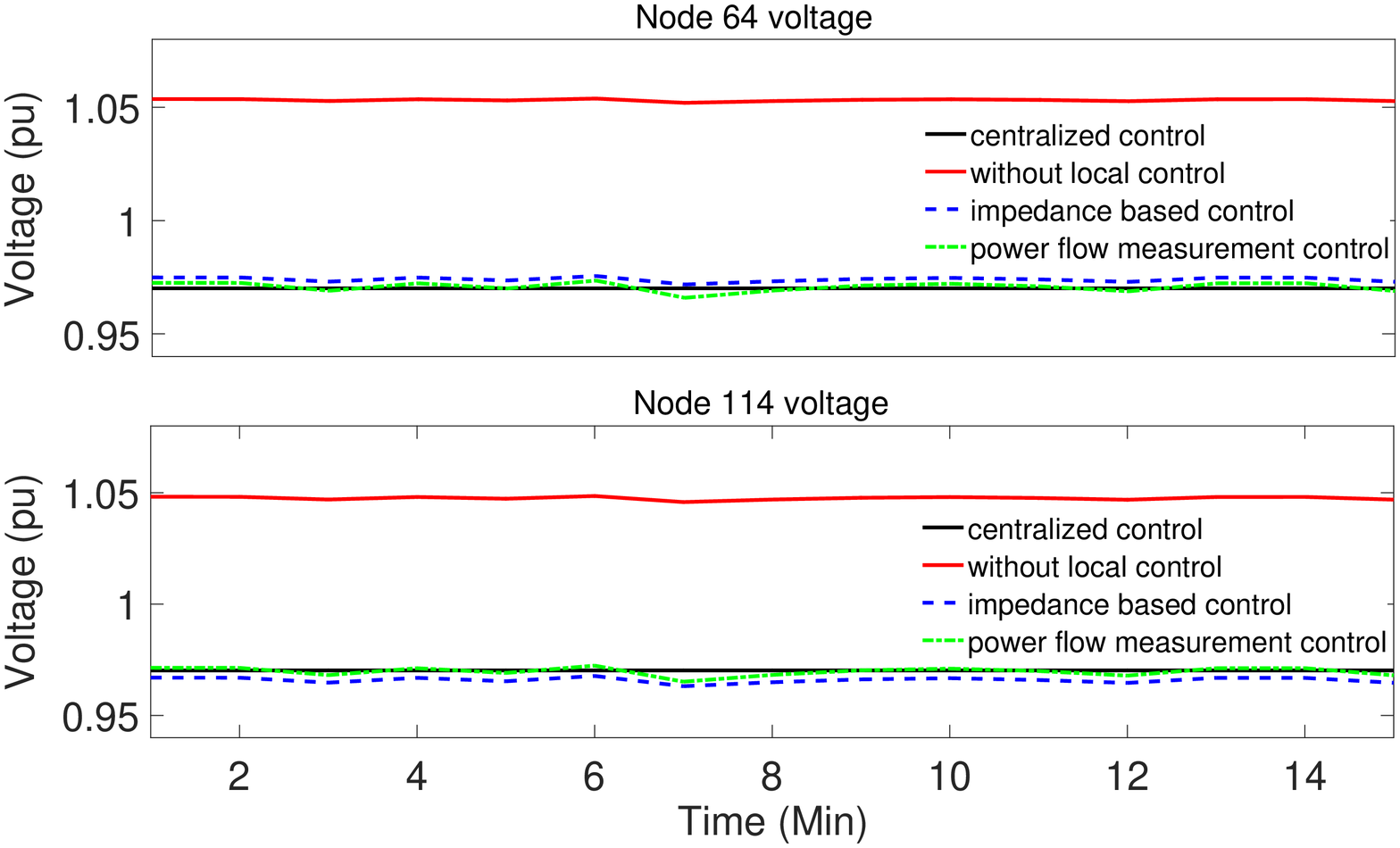}
\vspace{-0.4 cm}
	\caption{\footnotesize  Overvoltage scenario for the IEEE 123-bus system}
\vspace{-0.4 cm}
	\label{fig:12}
\end{figure}

\subsubsection{Mitigating Voltage Violations}
In this section, the overvoltage and undervoltage cases described in the previous section are observed in more detail. The overvoltage at nodes due to variable PV generation are observed when the load demand is less and the predicted PV power production is lower that the  actual PV power production. The Fig.\ref{fig:12}, shows the overvoltage scenario in the IEEE-123 node system at the node 64 and 114. In order to create this case, it is assumed that the predicted  PV power production is 10\% of the rated power production in the 15-min time interval and the actual power production for every 1-min (real-time) interval is around 70-80 \% of the rated value. The increased PV power will increase the voltage at the nodes. It can be observed from the figure that the voltage at nodes 64 and 114 are above 1.05 pu. This violates the specified ANSI voltage limits. The local control help to maintain the voltage close to the voltage obtained from centralized control. Both local control methods are able to mitigate the overvoltage violation and maintain the nodal voltages closer to the reference voltage obtained using the centralized controller. 

\begin{figure}[t]
\centering
\includegraphics[width=0.45\textwidth]{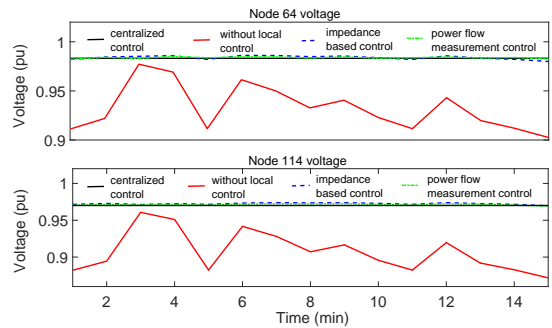}
\vspace{-0.3 cm}
	\caption{\footnotesize  Undervoltage  scenario for the IEEE 123-bus system}
\vspace{-0.5 cm}
	\label{fig:13}
\end{figure}

\begin{figure}[t]
\centering
\includegraphics[width=0.45\textwidth]{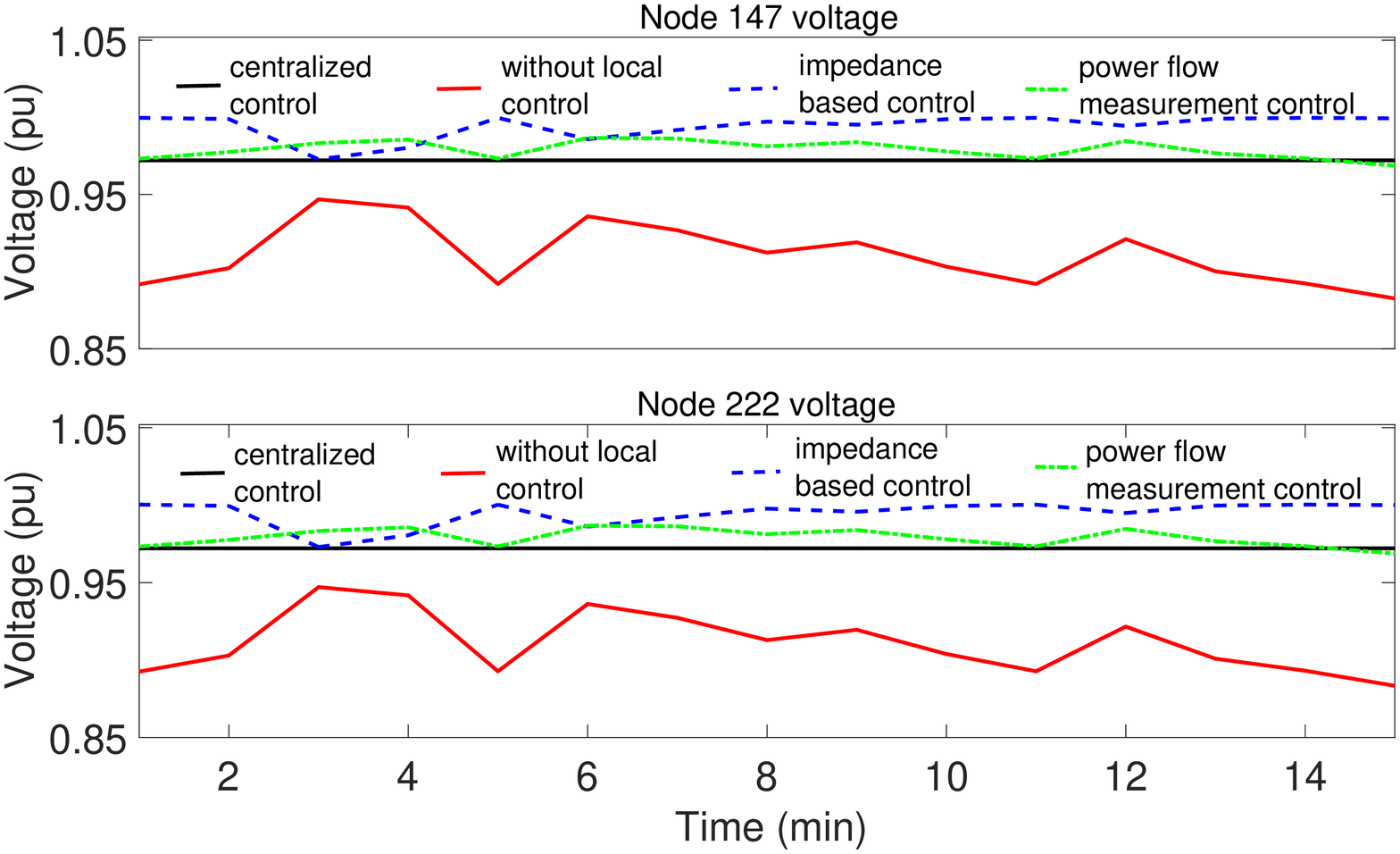}
\vspace{-0.3 cm}
	\caption{\footnotesize  Undervoltage  scenario for the  modified  R3-12.47-2 system}
\vspace{-0.3 cm}
	\label{fig:14}
\end{figure}

Similarly, the undervoltage condition at a node is observed when the load demand is high and the predicted PV power production is larger than the actual power production. In Fig.\ref{fig:13} and Fig.\ref{fig:14}, the undervoltage scenarios are simulated for the IEEE 123-bus and the  modified  R3-12.47-2 test feeders, respectively. The undervoltage scenario occur when the predicted PV power production is 90\% of the rated power production for the 15-min time interval but the actual power production (every 1-min interval or real-time) is around 10-20 \% of the rated power. For the IEEE-123 node system, node 64 and node 114 observe voltages below 0.95 pu (see Fig. \ref{fig:13}). The proposed local control is able to eliminate the undervoltage conditions at node 64 and 114 and maintain the voltage closer to the desired voltages obtained from centralized control. Similarly, the undervoltage scenario is created for the  modified  R3-12.47-2 system and it can observed from Fig.\ref{fig:14} that nodes 147 and 222 observe voltages less than 0.95 pu. The proposed local control is able to mitigate the undervoltage and maintain the voltage near to centralized voltage. Thus, local control methods are effective in mitigating potential voltage violations resulting from inaccurate forecast of PV generation that is used by the centralized controller. 

\subsection{Discussions}
In this section, we provide additional discussions regarding the utility of the proposed two-timescale controller. First, we provide the justification for adopting a central controller to regulate voltages at the MV level and not directly at the LV level. Note that to perform OPF, the central controller requires the knowledge of network connectivity, the locations and ratings of the legacy devices connected in the system, load demand, and DER generation. Unfortunately, at present,  behind-the-meter resources are not available to the network operator for the operational decision-making.  It requires the deployment of an extensive communication infrastructure to make the behind-the-meter data (of individual customers connected to the secondary feeders) accessible to the operator for operational decision-making. Therefore, we assume that aggregated load and DER measurements are available at the primary feeder (MV) nodes that the centralized controller uses for the operational decision making.

Next, we highlight the reason for having a two-timescale controller. The primary reason for adopting such architecture is to mitigate the voltage problems that the central controller is not able to handle. There are two reasons due to which the decisions of central controller alone may not be sufficient to mitigate the nodal voltage concerns at the LV level: (1) faster variability in generation (and/or load), (2) additional voltage drop in secondary feeders connecting MV and LV level. The local controller operates purely based on local node measurements i.e. sum of load and generation at a specific node. Also, the local voltages are affected by the net demand (generation) at the node. Hence, by acting based on local measurements, the local controller is able to act for the required voltage levels at the LV level. Thus, the LV voltage problems that the central controller is unable to mitigate are resolved by the local controllers. 

\section{Conclusion}
This paper presents a two-timescale control approach for system's voltage regulation devices to minimize the power consumption from voltage-dependent customer loads. In first stage, the centralized controller solves a linear three-phase OPF with mixed-integer/binary decision variables with the objective of minimizing the active power consumption from the substation. The centralized controller operates in every 15-min interval and sends the control signals to both legacy voltage control devices and the smart inverters. To address the concerns resulting from variable PV generation leading to nodal voltage fluctuations, at second stage, local control schemes coordinates with centralized controller is proposed. The local control is implemented for fast acting voltage control devices such as smart inverters; the control for legacy devices (capacitor banks and voltage regulators) is fixed for every 15-min interval based on optimal centralized controller decisions. The local control schemes are designed to obtain the required reactive power support from smart inverter to mitigate the voltage deviations resulting from variable PV generation. First, a Thevenin's impedance-based method is proposed to minimize the nodal voltage deviations (due to PV variability) wrt. the reference voltages obtained after implementing the centralized controller decisions. This method requires R/X ratio at the point-of-connection (POC) of PV and the change in local PV power generation wrt. the forecasted PV generation value used by the centralized controller. This approach, however, does not take the local decisions of other inverters into account. Therefore, to incorporate the effects of reactive power support from other smart inverters, a new method based on local power flow measurements is proposed. In addition to the change in local PV generation, this method also requires line flow measurements for the children nodes. 

The proposed two-timescale control approach is validated using the IEEE 123-bus and the  modified  R3-12.47-2 test systems. It is demonstrated that the proposed control approach is able to help realize CVR benefit for both feeders even for the cases with high PV variability. The local controls are able to reduce the voltage deviations caused due to variability in PV generation and maintain the nodal voltages closer to the reference voltage obtained from the centralized controller. The power flow measurement-based local control approach is shown to be relatively more effective compared to the Thevenin's impedance-based method. The MILP problem for the centralized controller takes on an average less than 30 sec. to solve for both the IEEE 123-bus (with a total of 267 single-phase nodes) and the  modified  R3-12.47-2 system (with a total of 860 single-phase nodes) test feeders. The local control is essentially instantaneous as it requires only simple arithmetic computations. Therefore, the proposed methods are suitable for the real-time control and operation of large-scale distribution feeders.   

	
\bibliographystyle{ieeetr}
\vspace{-0.1cm}
\bibliography{references}

\end{document}